\def\ANON{0} 
\def\PAGENUMS{0} 
\def\LINKBOXES{0} 
\def\ARXIV{1} 
\def\SUBMISSION{0} 
\def\AAM{0} 

\documentclass[conference]{IEEEtran}

\ifnum\SUBMISSION=1
    \def\ANON{0} 
    \def\PAGENUMS{0} 
    \def\LINKBOXES{0} 
\else
    \ifnum\ARXIV=1
        \def\AAM{1} 
        \def\LINKBOXES{0} 
        \def\PAGENUMS{1} 
        \def\ANON{0} 
    \fi
\fi

\ifnum\LINKBOXES=1
    \usepackage[breaklinks]{hyperref}
\else
    \ifnum\SUBMISSION=1
        \PassOptionsToPackage{bookmarks=false}{hyperref}
    \fi
    \usepackage[breaklinks,hidelinks]{hyperref}
\fi

\usepackage{graphicx}%
\usepackage{multirow}%
\usepackage{amsmath,amssymb,amsfonts}%
\usepackage{amsthm}%
\usepackage{mathrsfs}%
\usepackage{xcolor}%
\usepackage{textcomp}%
\usepackage{manyfoot}%
\usepackage{booktabs}%
\usepackage{algorithm}%
\usepackage{algorithmicx}%
\usepackage{algpseudocode}%
\usepackage{listings}%
\usepackage{orcidlink}
\usepackage{amsmath}
\usepackage{doi}
\usepackage{booktabs}
\usepackage{pifont} 
\newcommand{\cmark}{\ding{51}} 
\newcommand{\xmark}{\ding{55}} 
\usepackage{tablefootnote}
\usepackage{tabularx}
\usepackage{threeparttable}
\usepackage{balance}
\usepackage{siunitx}
\usepackage{comment}
\usepackage{subcaption}
\usepackage{fancyhdr} 
\usepackage{lastpage} 

\begin{document}


\title{Automated Risk Management Mechanisms in DeFi Lending Protocols: A Crosschain Comparative Analysis of Aave and Compound}

\ifnum\ANON=1
    \author{\IEEEauthorblockN{Anon}
    \IEEEauthorblockA{\\ \\ \\ \\}
    }
\else    
    \author{
        \IEEEauthorblockN{
            Erum Iftikhar\IEEEauthorrefmark{2}\IEEEauthorrefmark{1},
            Wei Wei\IEEEauthorrefmark{2} and 
            John Cartlidge\IEEEauthorrefmark{2}
            }
        \IEEEauthorblockA{\IEEEauthorrefmark{2}\textit{School of Engineering Mathematics and Technology, University of Bristol, Bristol, UK}\\
        \IEEEauthorblockA{\IEEEauthorrefmark{1}\textit{School of Advanced Studies, University of Camerino, Camerino, Italy}\\
        \texttt{erum.iftikhar@unicam.it, wei.wei@bristol.ac.uk, john.cartlidge@bristol.ac.uk}}}
    }
   
\fi

    
    
    
    
    
    

\maketitle

\ifnum\PAGENUMS=1
    \thispagestyle{fancy}
    \pagestyle{fancy}
    \fancyfoot[C]{\fontsize{9}{10} \selectfont Page \thepage ~of {\hypersetup{hidelinks}\pageref{LastPage}}}
    \ifnum\AAM=1 
        \fancyhead[C]{\fontsize{9}{10} \selectfont Accepted author manuscript: 7th Conf. on Blockchain Research \& Applications for Innovative Networks and Services (BRAINS 2025)}
    \else
        \renewcommand{\headrulewidth}{0pt} 
    \fi
\fi

\begin{abstract}
   Blockchain-based decentralised lending is a rapidly growing and evolving alternative to traditional lending, but it poses new risks. To mitigate these risks, lending protocols have integrated automated risk management tools into their smart contracts. However, the effectiveness of the latest risk management features introduced in the most recent versions of these lending protocols is understudied. To close this gap, we use a panel regression fixed effects model to empirically analyse the cross-version (v2 and v3) and cross-chain (L1 and L2) effectiveness of liquidation mechanisms, measured through TVL and total revenue as proxies for performance of the two most popular lending protocols, Aave and Compound, during the period Jan 2021 to Dec 2024. Our analysis reveals that liquidation events in v3 of both protocols lead to an increase in total value locked and total revenue, with stronger impact on the L2 blockchain compared to L1. In contrast, liquidations in v2 have an insignificant impact, which indicates that the most recent v3 protocols have better risk management than the earlier v2 protocols. We also show that L1 blockchains are the preferred choice among large investors for their robust liquidity and ecosystem depth, while L2 blockchains are more popular among retail investors for their lower fees and faster execution. 
\end{abstract}

\begin{IEEEkeywords}
    Blockchain, DeFi, Lending, Risk Management
\end{IEEEkeywords}

\section{Introduction}\label{sec:intro}
\begin{figure}[tb]
\centering
\fbox{%
\includegraphics[width=0.85\linewidth]{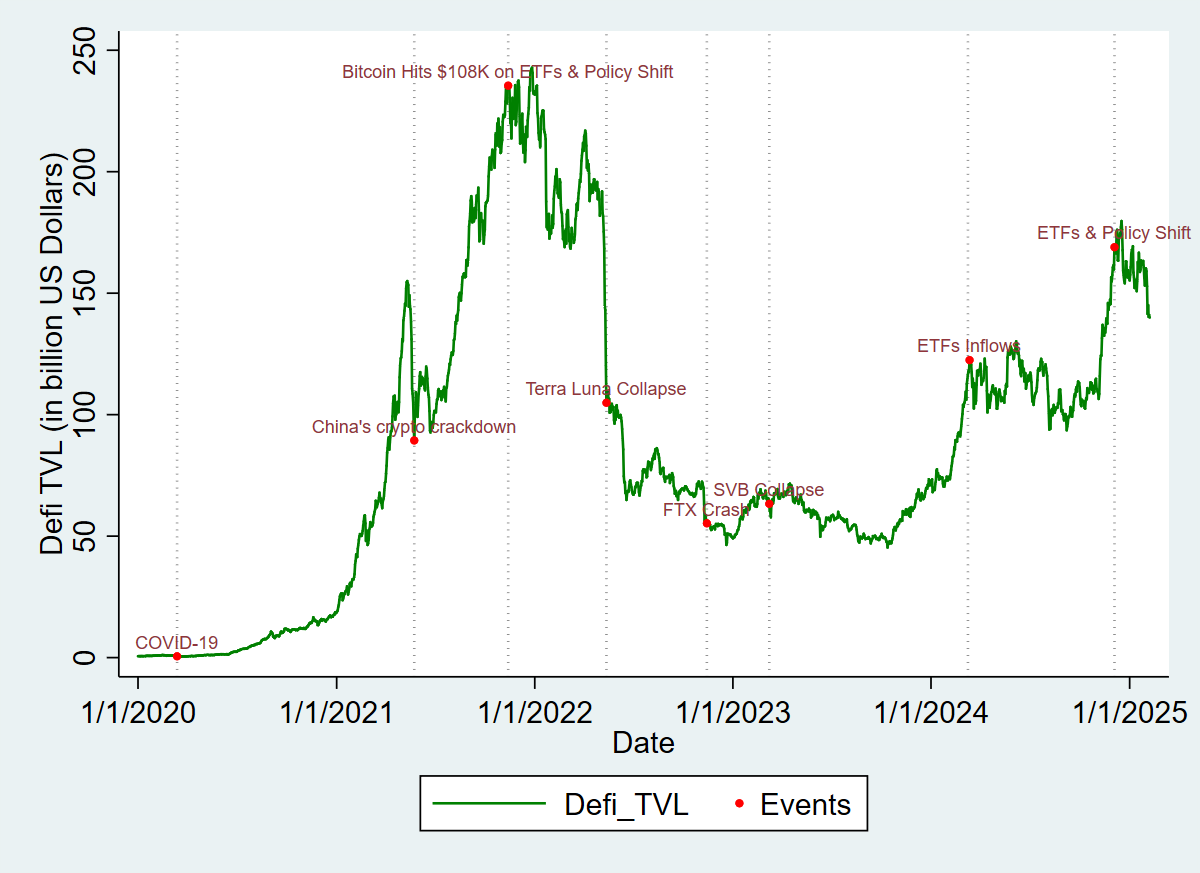}
}
\caption{\centering DeFi TVL 2020-2025 (Data Source: The Graph).}\label{fig:Defi_tvl}
\end{figure}

\noindent
Decentralised finance (DeFi) is a new and rapidly growing financial system built on blockchain technology and smart contracts. It offers more transparent, accessible, interoperable, secure, open, and peer-to-peer financial transactions \cite{kaplan2023blockchain} by replacing centralised intermediaries \cite{bertucci2024agents,sun2022liquidity}. These transactions are often more profitable than traditional financial systems \cite{sick2023economic} due to the low borrowing cost and high-return investments in crypto assets, even during a recession in traditional markets \cite{kuzenkov2023money}. Smart contracts can be programmed for a wide array of functions, replicating traditional financial services \cite{harvey2021defi} \cite{werner2022sok} such as lending, cryptocurrency exchange, asset management \cite{sun2022liquidity}, and insurance services \cite{huber2022risks}. The market size of DeFi, measured by the total value locked (TVL), has increased substantially over the last few years, peaking at USD \$179bn on 17 Dec 2024, as shown in Fig.~\ref{fig:Defi_tvl}. Among the various categories of DeFi protocols, lending is the largest, with nearly \$75bn TVL as of 31 Dec 2024 (\href{https://defillama.com/}{DeFiLlama}). 

However, despite its potential, DeFi faces significant risks, including regulatory concerns, scalability issues, liquidity risk, oracle risk, credit risk, and smart contract vulnerabilities \cite{huber2022risks}. The nascent ecosystem of DeFi also raises concerns about its stability and the lack of consumer protection mechanisms \cite{kuzenkov2023money}. These risks also manifest in DeFi lending markets, with the potential to cause considerable losses to participants and protocols. 

In DeFi lending, especially, the main risk is the liquidation risk \cite{qin2021empirical}, which occurs when the value of the collateral falls below the minimum collateralization threshold. Although liquidation risk primarily affects the borrower, it can also negatively affect lenders, liquidity providers, and protocols during black swan events of extreme market volatility \cite{gogol2024sok}. Analyzing the health and stability of lending protocols in response to liquidation risk is therefore crucial to understanding the automated risk management offered by different protocols, so that investors and policymakers can make informed decisions. 

To our knowledge, market data-driven research on liquidation risk management in DeFi lending protocols remains limited. In particular, we find that no previous studies have compared the performance and automated liquidity risk management of lending protocol versions v2 and v3, or the comparative effects of deployment on L1 and L2 blockchains. Of the available studies in the literature, their findings indicate that liquidation events can decrease the TVL of lending protocols \cite{sun2022liquidity,
luo2024piercing}; however, these works are limited to only earlier versions of the protocols (v1 and v2), which are deployed exclusively on L1 blockchains. Further investigation of risk management strategies used in the most recent version v3, deployed on both L1 and L2 blockchains, is still lacking.

In this work, we compare the top lending protocols, Aave and Compound, using recent data covering the period Jan 2021 to Dec 2024. We apply a panel regression fixed effects model to discover fresh insights into cross-version and cross-chain performance, as well as liquidation risk management. We use TVL and total revenue (TR) as proxies for protocol stability, regressed on liquidations-related and risk-related variables. 

{\bf Contribution:} Our findings reveal that v3 lending protocols have better risk management, particularly when deployed on L2 blockchains. Unlike the v2 protocols deployed on L1, where liquidations have no significant effect on revenue, v3 protocols---especially on L2---exhibit a positive relationship between liquidations and TVL, and a positive relationship between liquidations and TR. This suggests that the enhanced design of v3, combined with the scalability and lower transaction costs of L2, strengthens the resilience of the protocol during market stress. The results also point to a split in user preferences, such that institutional users gravitate toward L1 for its liquidity and ecosystem depth, while retail users prefer L2 for its cost efficiency.
 
The remainder of this study is arranged as follows: Section~\ref{sec:background} presents an introduction to the DeFi lending market and discusses the relevant literature in this field and how our research contributes to it. In Section~\ref{sec:data collection}, we present our data collection process. In Section~\ref{sec:Method-performance}, we discuss the empirical methodology used in our study. Section~\ref{sec:Results} interprets the relevant empirical findings and discusses the corresponding results. Finally, Section~\ref{Sec:conc} concludes. All code and data are publicly available.\footnote{\href{https://github.com/ErumIftikhar/DefiLending_Data_Code}{https://github.com/ErumIftikhar/DefiLending\_Data\_Code}.}

\section{Background}\label{sec:background}

\subsection{Evolution of DeFi lending Market, Protocols, and Advanced Risk Management}
\noindent
Decentralised lending refers to the practice of lending and borrowing facilitated directly through blockchain technology and smart contracts \cite{cornelli2024defi}, thereby bypassing traditional centralised intermediaries. According to \cite{heimbach2024defi}, decentralised lending platforms have emerged as a cornerstone, facilitating collateralised borrowing activities on an economically significant scale \cite{chiu2023fragility}, facilitating liquidity provision \cite{bhoyarcomparative}, and allowing users to earn interest on their crypto holdings or borrow funds.  

\begin{figure}[tb]
\centering
\fbox{%
\includegraphics[width=0.85\linewidth]{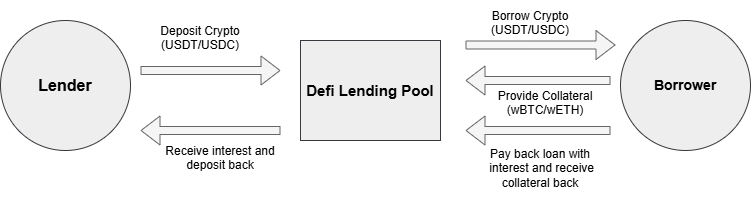}
}
\caption{\centering Defi lending mechanism.}\label{fig:Defi-lending}
\end{figure}

Lending and borrowing activities via DeFi lending protocols typically operate as shown in Fig.~\ref{fig:Defi-lending}. Lenders with excess funds provide assets to a lending smart contract. These assets are mostly in the form of stablecoins, e.g., USDT. Borrowers then provide collateral, mostly in the form of volatile coins, such as bitcoin, to borrow stablecoins. To access the service, borrowers pay a fee in the form of an interest payment, which is then passed on to lenders as a reward. The borrow rate of interest is set dynamically by a pre-defined function of crypto demand and supply \cite{bertucci2024agents}. This fluctuating rate protects the protocol against the run risk \cite{lehar2022systemic}. 

\begin{figure}[t]
\centering
\fbox{%
\includegraphics[width=0.85\linewidth]{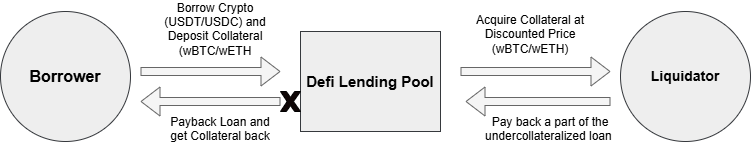}
}
\caption{\centering DeFi liquidation mechanism.}\label{fig:LiquidationMechanism}
\end{figure}

Due to the anonymity of borrowers and lenders and the volatile nature of the crypto market, the DeFi lending platform relies heavily on over-collateralised loans, which means that the amount of debt the borrowers can take on is typically lower than the value of the collateral deposited \cite{qin2021empirical,bertomeu2024measuring}.
The protocol ensures that enough collateral is locked over time and liquidation serves as a safety mechanism to reduce the risks of default \cite{mittal2023defi}. If the collateral depreciates below a {\em liquidation threshold}, the borrow position is designated as `under-collateralised' and the collateral can be liquidated by a third party liquidator. On such occasions, liquidators can claim a portion of the borrower's collateral at a discounted price---known as {\em liquidation bonus} for liquidators, and {\em liquidation penalty} for borrowers---to repay the debt, thus reducing the loan-to-value ratio (ratio of loan amount to collateral value) of the borrower’s loan, as shown in Fig.~\ref{fig:LiquidationMechanism}. However, despite the presence of liquidation mechanisms, defaults can still occur due to sudden drops in collateral value or insufficient incentives for liquidators to act \cite{bastankhah2024thinking}, which results in reduced capital efficiency and increased collateral illiquidity risk.

\begin{table*}[tb]
  \centering
  \caption{Feature Comparison: Aave and Compound.}
  \label{tab:features-comparison}
  \resizebox{1.0\linewidth}{!}{%
      \begin{tabular}{lccc ccc}
      \toprule
      \textbf{Feature} & \textbf{Aave v1} & \textbf{Aave v2} & \textbf{Aave v3} & \textbf{Compound v1} & \textbf{Compound v2} & \textbf{Compound v3}\\
      \midrule
      Year Launched & Jan 2020 & Dec 2020 & Mar 2022 & Sep 2018 & May 2019 & Aug 2022 \\
      Blockchain Layer & L1 & L1 & L1 \& L2 & L1 & L1 & L1 \& L2 \\
      Collateral Swaps & \xmark & \cmark & \cmark & \xmark & \xmark & Not supported directly\\
      Repayment with Collateral & \xmark & \cmark & \cmark & \xmark & \xmark & Not supported directly \\
      Flashloans & Introduced & Improved & Crosschain Flashloans & \xmark & \xmark & \xmark \\
      Isolation Mode & \xmark & \xmark & \cmark & \xmark & \xmark & Isolated pools \\
      Risk Param Customization & Static & Improved & Dynamic, asset-specific & Static & More flexibility & Asset-specific \\
      Efficiency Mode (eMode) & \xmark & \xmark & \cmark & \xmark & \xmark & \xmark \\
      Borrow/Supply Caps & \xmark & \xmark & \cmark & \xmark & \xmark & \cmark \\
      Oracle Source & Chainlink & Chainlink & Chainlink + CAPO & Centralized OPF & Chainlink & Chainlink + TWAP \\
      Liquidity Portals & \xmark & \xmark & \cmark & \xmark & \xmark & \xmark \\
      \bottomrule
      \end{tabular}
  }
\end{table*}

\begin{table*}[th]
  \centering
  \footnotesize
  \caption{Summary of prior findings on the impact of liquidations on TVL and revenue in DeFi lending protocols.}
  \label{tab:prior-work}
  \resizebox{1.0\linewidth}{!}{%
  \begin{tabular}{@{}c p{7.5cm} c p{5cm} c@{}}
    \toprule
    \textbf{Ref} & \textbf{Finding} &  \textbf{Timeframe} & \textbf{Protocols} & \textbf{Blockchain} \\
    \midrule
    \cite{sun2022liquidity} & Liquidations cause a decrease in TVL & Dec 2019 -- Jan 2023 & Aave v1, Aave v2, Compound v3 & Layer 1 \\
    \addlinespace    
    \cite{luo2024piercing} & Liquidations cause a decrease in TVL & Jan 2021 -- Mar 2024 & Aave v2, MakerDAO, JustLend & Layer 1  \\
    \addlinespace
    \cite{sun2022liquidity} & Insignificant relationship between liquidations and total revenue & Dec 2019 -- Jan 2023 & Aave v1, Aave v2, Compound v3 & Layer 1  \\
   \addlinespace
    \cite{sinyugin2023decentralized} & Mass liquidations can sometimes signal efficient risk management & Descriptive analysis & Aave and Compound & Not specified \\
    \bottomrule
  \end{tabular}
  }
  \end{table*}

Table~\ref{tab:features-comparison} summarizes the functionality of Aave and Compound, two of the oldest and most successful DeFi lending platforms. Aave has the highest TVL in the DeFi lending space and is a decentralised open-source platform operating on 11 different blockchains. Compound finance is also among the top five lending protocols and operates on nine different blockchains. Notably, initial versions (i.e., v1, v2) of these protocols only implemented basic risk controls, such as collateral factors and static risk parameters (liquidation threshold, liquidation penalties, borrow/lending rates), whereas upgraded versions (i.e., v3) introduced more sophisticated tools that are explicitly designed to mitigate systemic vulnerabilities. These tools differ from traditional risk management practices (ring-fencing, credit exposure limits, and portfolio rebalancing) by being automated, transparent, and enforced via smart contracts. The dynamic risk parameters and targeted mechanisms provide more granular control over asset-specific risk and reduce the likelihood of systemic contagion.   These upgraded versions are launched on both L1 and L2 blockchains: L1 blockchains (e.g., Ethereum) denote the base networks that establish the primary consensus processes to achieve decentralization and guarantee primary on-chain activities; while L2 blockchains (e.g., Arbitrum, Optimism) inherit security and decentralization from L1 and increase transaction throughput, offering low transaction fees and faster transactions \cite{song2024advancing}.

Risk management in lending protocols relies on automated mechanisms like over-collateralisation ratios, governance frameworks, and liquidation triggers to protect participants and maintain stability. However, these measures can be insufficient during periods of high volatility, market congestion, liquidation spirals, and systemic risk. High transaction costs can also reduce their effectiveness in maintaining stability \cite{lehar2022systemic,bertomeu2024measuring}. The latest v3 versions of the protocols address these shortcomings by incorporating advanced risk management features, as shown in Table~\ref{tab:features-comparison}; including enhanced liquidation incentives and dynamic collateral requirements, which allow users to adjust their borrowing parameters to match their risk profile, thus mitigating liquidation cascades and improving resilience. Isolation mode prevents contagion from volatile assets; efficiency mode, optimizes capital utilization for correlated assets; and caps on borrow and supply asset limits, overexposure to risky tokens. Furthermore, the integration of L2 blockchain with v3 protocols has reduced transaction cost and increased throughput, thus facilitating faster and more efficient collateral modifications and immediate liquidation triggers, which mitigates the prospect of delayed or unsuccessful liquidations. As more than 70\% of liquidations occur in the absence of any price decline of crypto assets \cite{moallemi2024analysis}, the latest protocols also observe market sentiment and patterns for efficient risk management. These innovations collectively facilitate a more resilient and efficient risk management system (particularly when compared with v2 on L1 blockchain). Our study focuses on investigating those risk parameters that exert a direct impact on the liquidation risk mechanism, such as to loan-to-value ratio, liquidation penalty, and liquidation threshold.

\subsection{Existing work on liquidation risks in DeFi lending}
\noindent
Here, we review the relatively limited literature on the evaluation of liquidation risks in DeFi lending protocols. 

Study \cite{qin2021empirical} investigated liquidations in lending protocols by considering the potential to manipulate the existing system and suggested an alternative liquidation mechanism called a ``reversible call option''. Study \cite{warmuz2022toxic} also investigated liquidation spirals and recommended changes to existing liquidation mechanisms. In \cite{lehar2022systemic}, an inherent systemic fragility was observed in DeFi lending markets, resulting from liquidators selling the collateral during a price drop, leading to a further drop in prices and causing cascading liquidations of other loans. Related to systemic issues, \cite{cohen2023paradox} highlighted that any incentive for liquidators to act when a loan is in distress would be associated with a subsequent incentive to manipulate prices and cause liquidations of loans. 

Previous studies, with a particular focus on the relationship between liquidations and TVL, highlight the {\em negative impact} of liquidations on the TVL of the protocols. Study \cite{luo2024piercing} investigated the impact of ETH volatility on TVL of the protocol via liquidations. They performed an empirical and sensitivity analysis using the decline in ETH as a price shock to TVL, while controlling different factors, such as the gas price, the liquidation threshold, and VIX. Their findings showed that TVL is highly unstable during market downturns due to cascading liquidation events. The impact of different risk indicators on the performance of the protocol was also investigated by \cite{sun2022liquidity}. They used different performance metrics including TVL and TR and confirmed the negative impact of liquidations on TVL, but found insignificant results for the impact on TR. The key findings of  \cite{sun2022liquidity,luo2024piercing} are summarised in Table~\ref{tab:prior-work}. Note that these studies focused mainly on earlier versions, that is v1 and v2 of the protocols; only \cite{sun2022liquidity} included v3 of the Compound protocol, but within a relatively limited time frame (from protocol launch in Aug 2022, until Jan 2023).

Despite a growing body of literature on DeFi risks, existing studies have largely overlooked the effectiveness of evolving risk management mechanisms within lending protocols, particularly in the most recent v3 versions launched on L1 and L2 blockchains. Our study adds to the growing literature that focuses on the efficiency of risk management system in upgraded versions of lending protocols by investigating protocol's advanced features, transaction costs, volatility, and sentiment in crypto and traditional markets.

\section{Data Collection}\label{sec:data collection}
\begin{figure}[tb]
\centering
\fbox{%
\includegraphics[width=0.85\linewidth]{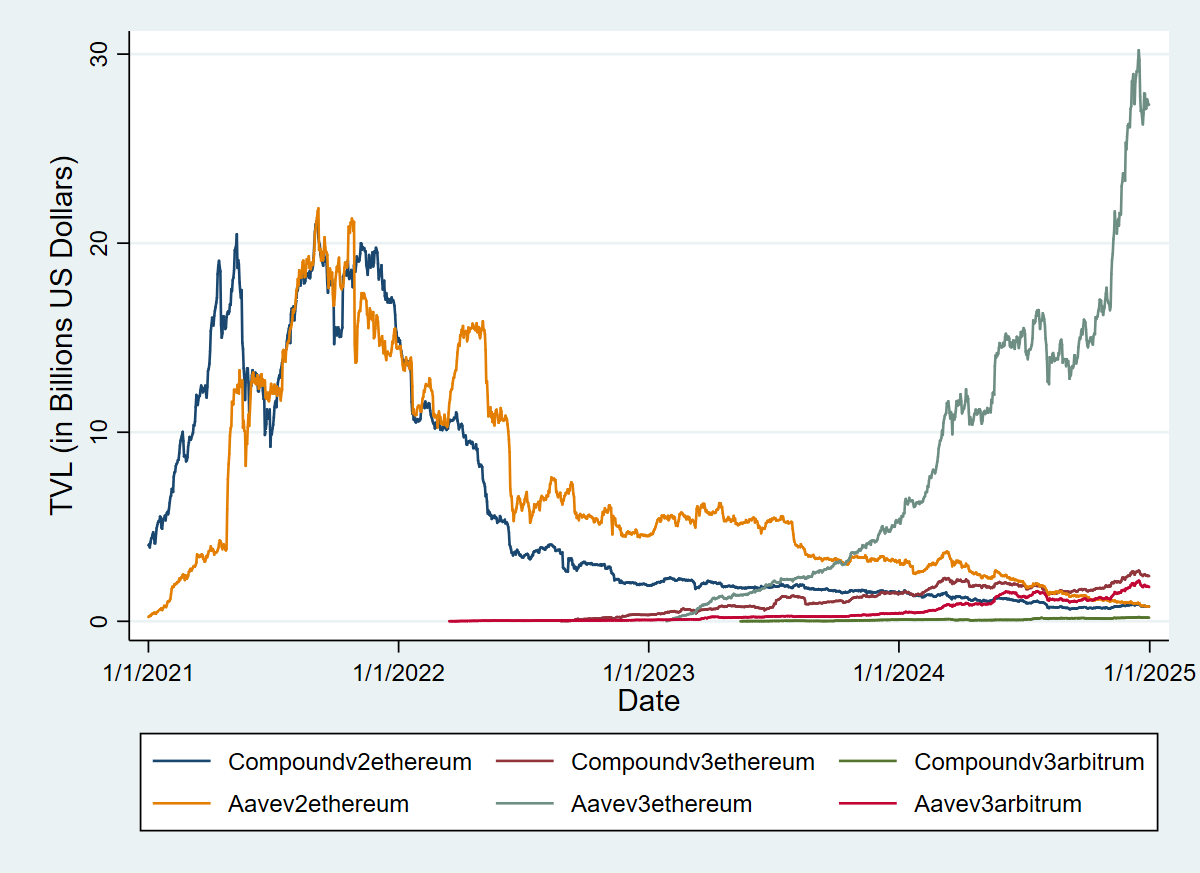}
}
\caption{\centering Crosschain TVL of Aave and Compound.}\label{fig:TVL}
\end{figure}

\noindent
We focus on Aave and Compound in analyzing the performance of risk management in lending protocols. The period of the research ranges from 01 Jan 2021 to 31 Dec 2024, which includes significant events of boom and crises, aligning with our research goals (see Fig.~\ref{fig:Defi_tvl}). For cross-version comparison, v2 and v3 versions are selected (see Table~\ref{tab:features-comparison} for feature descriptions), while v1 was omitted due to its deprecated status. In conducting a cross-chain analysis, we focus on Ethereum as a representative of Layer 1 blockchains and Arbitrum for Layer 2 blockchains, chosen based on their leading TVL during the time period under consideration. The TVL data for the selected protocols is presented in Fig.~\ref{fig:TVL}. 

To include data related to the performances of protocols, we collect the underlying asset data—specifically, the daily adjusted closing prices of ETH—from \href{www.coinmarketcap.com}{CoinMarketCap}. Core protocol-level data, including TVL, TR, liquidation volume, withdrawal volume, loan-to-value ratios, liquidation penalties, and liquidation thresholds for each protocol, are obtained from \href{https://thegraph.com/}{The Graph}. Additionally, borrowing rates for USDC in Aave and Compound are sourced from \href{https://aavescan.com/}{Aavescan}. We also collect gas price data for the Ethereum and Arbitrum blockchains from \href{https://etherscan.io/chart/gasprice }{Etherscan} and \href{https://arbiscan.io/chart/gasprice}{Arbiscan}, respectively, to account for the transaction costs of each layer.

To capture market sentiment, we include the fear and greed index (FGI)  of the crypto market,\footnote{FGI is an indicator between `0'  (extreme fear) and `100' (extreme greed), implying investors' selling and buying tendency.} and CBOE volatility index (VIX) from \href{https://alternative.me/crypto/fear-and-greed-index/} {alternative.me} and \href{https://www.lseg.com/en/data-analytics/financial-data/workspace-datasets}{LSEG workspace}, respectively. We also define a risk metric in Equation~\eqref{eq:risk}, which we call the {\em Default Risk Metric} (DRM). A high value of the DRM suggests that the protocol employs a stringent risk management policy, resulting in a higher likelihood of borrower liquidations. Conversely, a low DRM indicates a more lenient risk framework, offering a more accommodating environment for borrowers.

\begin{equation}\label{eq:risk}
        \text{DRM} = 
        \frac{\text{Loan-To-Value}}{\text{Liq. Threshold}}  \;   
        \left(1 + \text{Liq. Penalty} \right)
\end{equation}

\vspace{5mm}
The final dataset used for further analysis comprises 1456 observations, following the removal of extreme outliers. Summary statistics for these data are reported in Table~\ref{tab:SummaryStats}. 

\begin{table}[tb]
\centering
\caption{Summary Statistics}
\label{tab:SummaryStats}
\begin{tabularx}{0.9\linewidth}{l *{3}{S[table-format=+.2]} S[table-format=+2.2]}
    \toprule
    \textbf{Variable} & \textbf{Mean} & \textbf{Std} & \textbf{Min} & \textbf{Max} \\
    \midrule
    ETH Returns & 0.00 & 0.04 & -0.31 & 0.22 \\ 
    TVL (\$bn) & 4.46 & 5.54 & 0.01 & 30.21 \\ 
    Total Revenue (\$bn) & 0.13 & 5.09 & 0.00 & 260.00\\ 
    Liquidations (\$mn) & 0.47 & 5.16 & 0.00 & 232.22 \\
    Withdrawals (\$bn) & 0.31 & 0.88 & 0.00 & 25.40 \\
    USDC Borrow APR & 5.76 & 7.11 & 0.71 & 63.57 \\ 
    Gas Ethereum (ETH) & 41.52 & 52.06 & 0.00 & 939.59 \\ 
    Gas Arbitrum (ETH) & 3.32 & 1.43 & 0.00 & 4.78 \\ 
    FGI & 51.12 & 20.75 & 6.00 & 95.00 \\ 
    VIX & 18.44 & 5.17 & 11.86 & 38.57 \\ 
    DRM & 7.19 & 4.92 & 0.95 & 11.10 \\ 
\bottomrule
\end{tabularx}

\begin{minipage}{0.9\linewidth}
    \footnotesize
    \vspace{1mm}
    Note: Summary of all variables included in our data from 01 Jan 2021 to 31 Dec 2024, with outliers removed. Raw data consists of daily frequency of 1461 observations. Total revenue had some extreme outliers, driven by large scale liquidations in Aave v3, which generated substantial liquidation fees. These daily observations were removed (5 in total) using a $1\%$ cut-off, resulting in 1456 observations. However, all findings reported in this study are unchanged when analysis is performed on the full data, including outliers.
\end{minipage}
\end{table}


\section{Methodology}\label{sec:Method-performance}
\noindent
Regression analysis is the most common method to determine the relationship between dependent and independent variables in DeFi markets. Studies that have used regression analysis to assess various aspects and risks in DeFi lending markets include \cite{bertucci2024agents, sun2022liquidity, cornelli2024defi, lehar2022systemic, bertomeu2024measuring, bastankhah2024thinking, moallemi2024analysis, abraham2024crypto}. 
In this study, we employ a panel regression model with fixed effects to quantify how liquidations---driven by various risk factors---affect the performance (proxied by TVL and TR) of DeFi lending protocols. As our dataset spans multiple protocols (Aave and Compound, across v2/v3 and L1/L2 deployments) over time, panel regression analysis is well-suited because it captures both the cross-sectional variation between protocols and the time-series variation within each protocol. Panel regression with fixed effects controls for unobserved, time-invariant heterogeneity and improves estimation efficiency by leveraging repeated observations over time \cite{wooldridge2010econometric}. The regression model is illustrated in Equation~\eqref{eq:regression}.

\begin{align}
\label{eq:regression}
\mathrm{Perf}_{i,t}    = & \, \alpha_i
+ \beta_1 \mathrm{Liq}_{i,t}
+  \beta_2 \mathrm{Liq}_{i,t} \mathrm{Lay}_{i}
+  \beta_3 \mathrm{Liq}_{i,t} \mathrm{Ver}_{i} \nonumber \\
& +\beta_4 \mathrm{Liq}_{i,t} \mathrm{Platf}_{i}
+ \gamma_1 \mathrm{DRM}_{i,t}
+ \gamma_2  \mathrm{VolETH}_{t} \nonumber\\
&+ \gamma_3  \mathrm{VolGasPri}_{i,t}
+\gamma_4 \mathrm{VolUSDCbor}_{i,t}\nonumber \\
&+ \gamma_5 \mathrm{Wd}_{i,t} 
+ \gamma_6 \mathrm{FGI}_{t}
+ \gamma_7 \mathrm{VIX}_{t} 
+  \varepsilon_{i,t} 
\end{align}

\noindent 

The dependent variable, $\text{Perf}_{i,t}$, measures the performance of protocol $i$ at time $t$, using either TVL or TR data after logarithmic transformation. Our main explanatory variable is the logged liquidation volume, $\mathrm{Liq}_{i,t}$. We include several control variables to account for potential confounding factors. The variable $\text{DRM}_{i,t}$ captures the level of default risk associated with protocol $i$ at time $t$. Market volatility is represented by $\text{VolETH}_{i,t}$ for the collateral asset ETH, $\text{VolGasPri}_{i,t}$ for the volatility of gas prices, and $\text{VolUSDCbor}_{i,t}$ for the volatility of USDC borrow rate. The volatilities are 7-day rolling standard deviation of the log-returns. Protocol-level activity is measured by $\text{Wd}{i,t}$, which is the logged total withdrawal volume from protocol $i$ at time $t$. To control for broader market sentiment, we incorporate $\text{FGI}_{t}$ as a proxy for investor sentiment in the cryptocurrency market and $\text{VIX}_{t}$ for sentiment in the traditional financial market. The sentiment variables are used after logarithmic transformation.

To capture heterogeneity across blockchain layers, protocol versions, and lending platforms, we introduce three dummy variables. To indicate the blockchain layer, we use $\mathrm{Lay}_i\in\{0=\text{L1}, 1= \text{L2}\}$; for the protocol version, we use $\mathrm{Ver}_i\in\{0=\text{v2}, 1=\text{v}3\}$; and for the individual protocol, we use $\mathrm{Platf}_i\in\{0=\text{Compound},1=\text{Aave}\}$. In  cross-version and cross-chain analysis, we include the dummy variable $\text{Ver}_i$ and $\text{Lay}_i$ in the regression models---referred to as Model 1 (regression on TVL) and Model 2 (regression on TR). For the analysis of cross-protocol and cross-version effects on performance, we incorporate the dummy variables $\text{Ver}_i$ and $\text{Platf}_{i}$, corresponding to Model 3 (regression on TVL) and Model 4 (regression on TR). A comprehensive regression model including all variables is presented in Appendix~\ref{app:FullModel}.

Before conducting the regression analysis, the multicollinearity was assessed using the Variance Inflation Factor (VIF). The VIF values ranged from 1.1 to 2.1, well below the commonly used threshold of 3, indicating that there is no significant multicollinearity among the explanatory variables (see Appendix~\ref{app:multicoll} for details).

\section{Results and Findings}\label{sec:Results}

\begin{table}[tb]
\centering
\caption{Cross-Version, Cross-Chain, and Cross-Protocol Regression of TVL and TR on Liquidations}
\label{tab:MergedTables}
\begin{tabularx}{0.95\linewidth}{l *{4}{S[table-format=+1.2, table-space-text-post=***]}}
\toprule
\textbf{Variable} & \textbf{Model 1} & \textbf{Model 2} & \textbf{Model 3} & \textbf{Model 4} \\
\midrule
 Liq & -0.01~$^{***}$ & 0.00 & 0.00 & 0.00 \\
Liq:Lay & 0.05~$^{***}$ & 0.09~$^{***}$ &   &      \\
 Liq:Ver & 0.03~$^{***}$ & 0.02~$^{*}$ & 0.05~$^{***}$ & 0.06~$^{***}$ \\
 Liq:Platf &       &       & -0.01~$^{*}$ & 0.00 \\
 DRM & 0.01~$^{***}$ & 0.01~$^{***}$ & 0.01~$^{***}$ & 0.01~$^{***}$ \\
 VolETH & -1.90~$^{***}$ & -2.23~$^{**}$ & -1.94~$^{***}$ & -2.25~$^{**}$ \\
  VolGasPri &-0.15~$^{***}$ & -0.19~$^{***}$ & -0.15~$^{***}$ & -0.19~$^{***}$ \\
 VolUSDCbor &  0.02~$^{***}$ & 0.05~$^{***}$ & 0.02~$^{***}$ & 0.05~$^{***}$ \\ 
Wd & 0.40~$^{***}$ & 0.49~$^{***}$ & 0.40~$^{***}$ & 0.49~$^{***}$ \\
 FGI & 0.08~$^{***}$ & 0.46~$^{***}$ & 0.09~$^{**}$ & 0.48~$^{***}$ \\
 VIX & -0.27~$^{***}$ & -0.76~$^{***}$ & -0.29~$^{***}$ & -0.79~$^{***}$ \\
 Adj R-sq & 0.61 & 0.57 & 0.61 & 0.56 \\
\bottomrule

\end{tabularx}
\begin{minipage}{0.95\linewidth}
\footnotesize
\setlength{\baselineskip}{7pt} 
\vspace{1mm}
Note: This table includes estimates of four panel regression models performed on cross-chain, cross-version TVL (Model~1) and TR (Model~2) as dependent variables, regressed against liquidations and relevant control variables. In Model~3 and Model~4, regression estimates of cross-protocol, cross-version analysis are reported, where TVL (Model~3) and TR (Model~4) as dependent variables are regressed against liquidations and relevant control variables. All models are estimated as follows:
\\

Model~1: \(
\mathrm{TVL}_{i,t} = \alpha_i
+ \beta_1 \mathrm{Liq}_{i,t}
+  \beta_2 \mathrm{Liq}_{i,t} \mathrm{Lay}_{i}
+  \beta_3 \mathrm{Liq}_{i,t} \mathrm{Ver}_{i}
+ \gamma_1 \mathrm{DRM}_{i,t}
+\gamma_2 \mathrm{VolUSDCbor}_{i,t}
+ \gamma_3  \mathrm{VolGasPri}_{i,t}
+ \gamma_4  \mathrm{VolETH}_{t} 
+ \gamma_5 \mathrm{FGI}_{t} 
+ \gamma_6 \mathrm{VIX}_{t} 
+ \gamma_7 \mathrm{Wd}_{i,t} 
+  \varepsilon_{i,t}  \).\\

Model~2: \(
\mathrm{TR}_{i,t} = \alpha_i
+ \beta_1 \mathrm{Liq}_{i,t}
 +\beta_2 \mathrm{Liq}_{i,t} \mathrm{Lay}_{i} 
+  \beta_3 \mathrm{Liq}_{i,t} \mathrm{Ver}_{i}
+ \gamma_1 \mathrm{DRM}_{i,t}
+\gamma_2 \mathrm{VolUSDCbor}_{i,t}
+ \gamma_3  \mathrm{VolGasPri}_{i,t}
+ \gamma_4  \mathrm{VolETH}_{t} 
+ \gamma_5 \mathrm{FGI}_{t} 
+ \gamma_6 \mathrm{VIX}_{t} 
+ \gamma_7 \mathrm{Wd}_{i,t} 
+  \varepsilon_{i,t}  \).\\

Model~3: \( 
\mathrm{TVL}_{i,t}    = \, \alpha_i
+ \beta_1 \mathrm{Liq}_{i,t}
+  \beta_3 \mathrm{Liq}_{i,t} \mathrm{Ver}_{i} 
 +\beta_4 \mathrm{Liq}_{i,t} \mathrm{Platf}_{i}
+ \gamma_1 \mathrm{DRM}_{i,t}
+ \gamma_2  \mathrm{VolETH}_{t} 
+ \gamma_3  \mathrm{VolGasPri}_{i,t}
+\gamma_4 \mathrm{VolUSDCbor}_{i,t}
+ \gamma_5 \mathrm{Wd}_{i,t} 
+ \gamma_6 \mathrm{FGI}_{t}+ \gamma_7 \mathrm{VIX}_{t} 
+  \varepsilon_{i,t} \).\\

Model~4: \( 
\mathrm{TR}_{i,t}    = \, \alpha_i
+ \beta_1 \mathrm{Liq}_{i,t}
+  \beta_3
\mathrm{Liq}_{i,t} \mathrm{Ver}_{i} 
 +\beta_4 \mathrm{Liq}_{i,t} \mathrm{Platf}_{i}
+ \gamma_1 \mathrm{DRM}_{i,t}
+ \gamma_2  \mathrm{VolETH}_{t} 
+ \gamma_3  \mathrm{VolGasPri}_{i,t}
+\gamma_4 \mathrm{VolUSDCbor}_{i,t}
+ \gamma_5 \mathrm{Wd}_{i,t} 
+ \gamma_6 \mathrm{FGI}_{t} + \gamma_7 \mathrm{VIX}_{t} 
+  \varepsilon_{i,t} \).\\

Asterisks indicate statistical significance of estimates; with $^{*}$ indicating $p < 0.10$, $^{**}$ indicating $p < 0.05$, and $^{***}$ indicating $p < 0.01$.

\end{minipage}

\end{table}

\subsection{Cross-version and cross-chain analysis of liquidations}
\noindent
Model 1 in Table~\ref{tab:MergedTables} presents TVL regression results for lending protocols v2 and v3 across Ethereum (L1) and Arbitrum (L2) chains. For v2 lending protocols deployed on the Ethereum blockchain, we find that a $1\%$ increase in liquidations is associated with a $0.01\%$ {\em decrease} in TVL for these protocols (i.e., $\mathrm{Liq} = -0.01$). This negative relationship observed in v2 protocols is consistent with previous findings \cite{sun2022liquidity,luo2024piercing} (also see Table~\ref{tab:prior-work}). In contrast, we find that TVL in v3 protocols deployed on L1 and L2 chains has a {\em positive} significant relationship with liquidation, with a $1\%$ increase in liquidations associated with a $0.02\%$ {\em increase} in TVL on L1 ($\mathrm{Liq} + \mathrm{Liq\mathord{:}Ver}= (-0.01) + 0.03 = 0.02$), and a $0.07\%$ {\em increase} in TVL on L2 ($\mathrm{Liq} + \mathrm{Liq\mathord{:}Lay}+ \mathrm{Liq\mathord{:}Ver} = (-0.01) + 0.05 + 0.03 = 0.07$). The overall {\em positive} relationship on v3 protocols---i.e., an increase in liquidations leads to an {\em increase} in TVL---is a new finding that has not been reported before.

\begin{figure*}[t]
    \centering
    \begin{subfigure}[t]{0.33\textwidth}
        \centering
        \includegraphics[width=\textwidth]{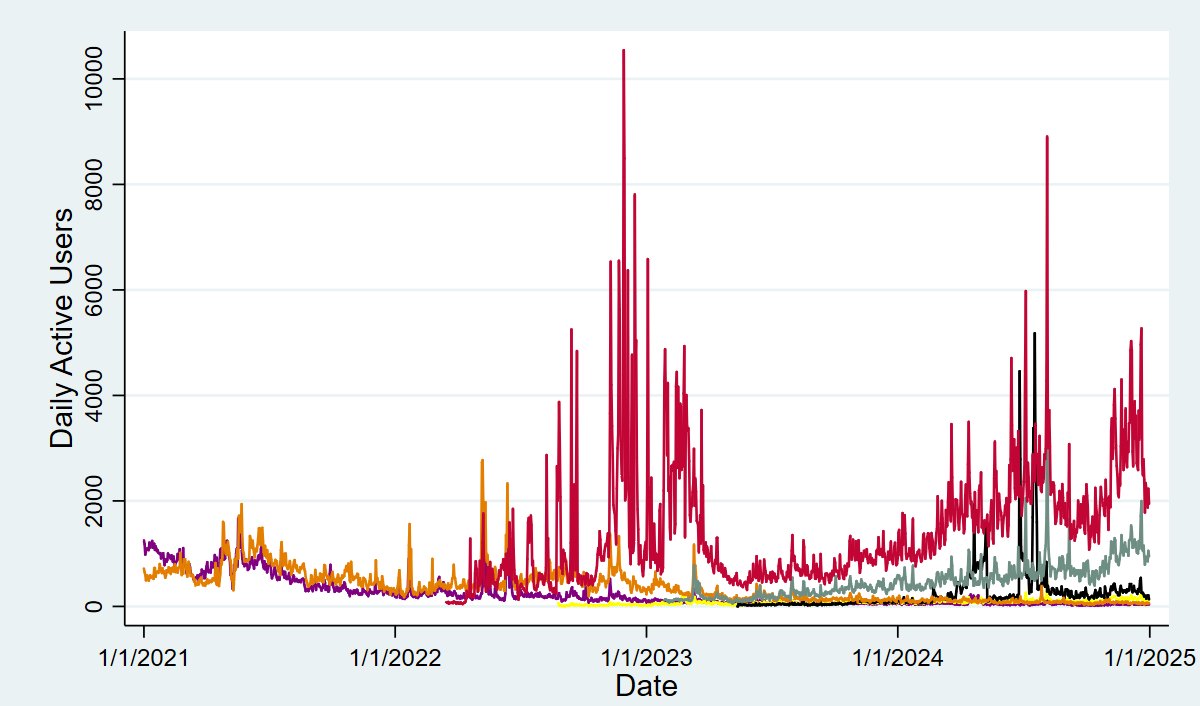}
        \caption{Daily active users.}\label{fig:Users}
    \end{subfigure}%
    ~ 
    \begin{subfigure}[t]{0.32\textwidth}
        \centering
        \includegraphics[width=\textwidth]{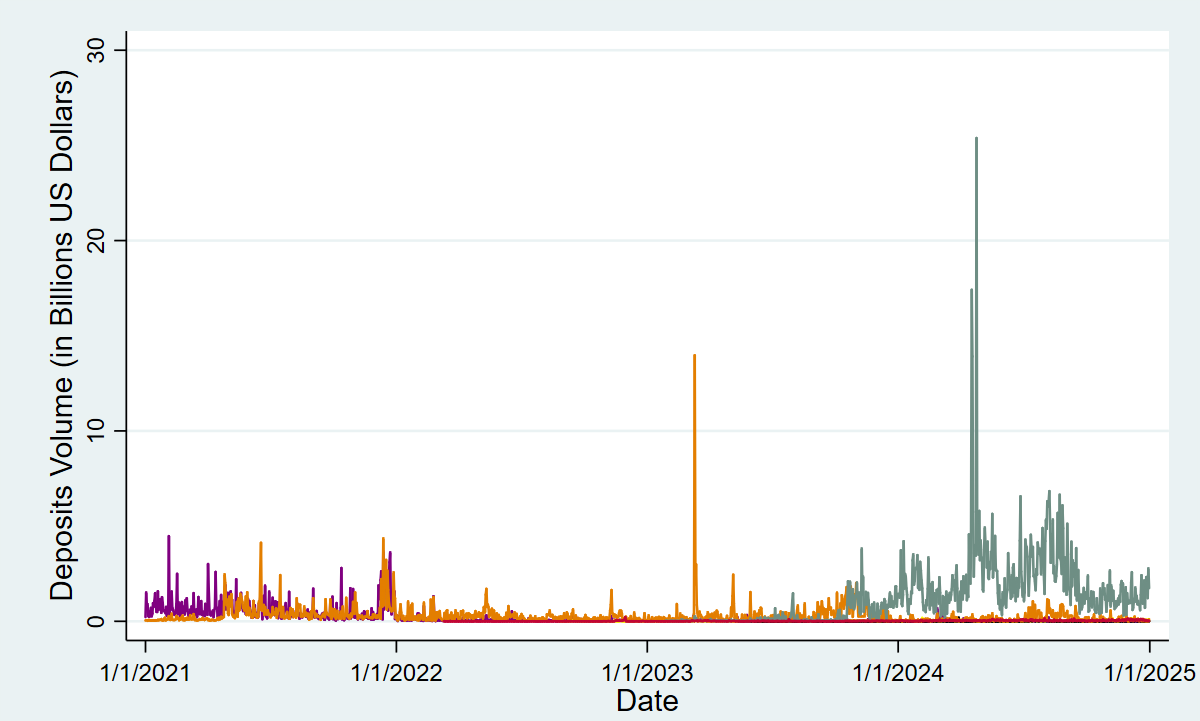}
        \caption{Deposit volume.}\label{fig:Deposits}
    \end{subfigure}%
    ~ 
    \begin{subfigure}[t]{0.33\textwidth}
        \centering
        \includegraphics[width=\textwidth]{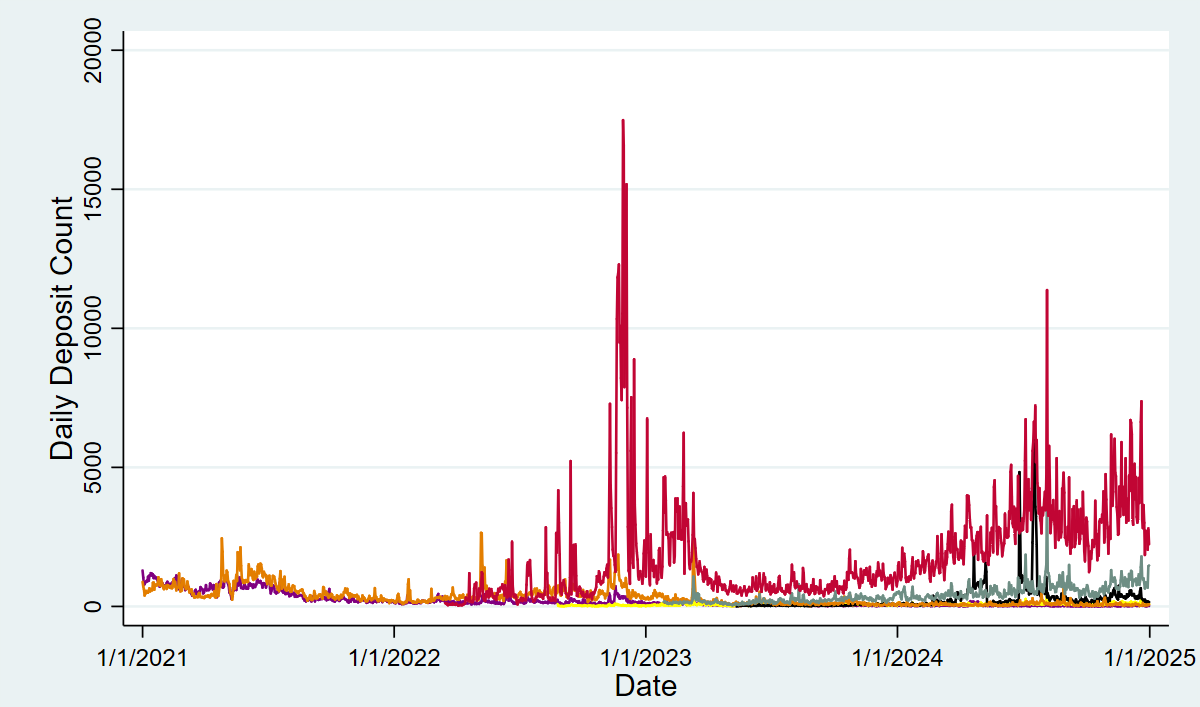}
        \caption{Daily deposit count.}\label{fig:DailyDeposits}
    \end{subfigure}\\[1ex]
    \begin{subfigure}[t]{\textwidth}
        \centering
        \includegraphics[width=0.4\textwidth]{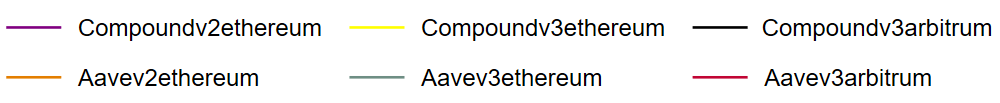}
    \end{subfigure}\\[1ex]    
    \caption{Usage comparison of lending protocols.}\label{fig:usage}
\end{figure*}

It should be noted that the coefficient $\mathrm{Liq\mathord{:}Lay}$ is significantly {\em positive} at $0.05$, indicating that liquidation has a heterogeneous impact on TVL across different layers under the same version, specifically in v3. This suggests that a $1\%$ increase in liquidation on Arbitrum is associated with a $0.05\%$ {\em greater increase} in TVL compared to Ethereum, highlighting a differentiated response to liquidation activity between the two layers. This is a new finding that has not been reported before.

Model 2 in Table~\ref{tab:MergedTables} presents the regression on TR for protocols v2 and v3 across Ethereum (L1) and Arbitrum (L2) chains. An insignificant coefficient for $\mathrm{Liq}$ was found, indicating an insignificant impact of liquidation on TR for the v2 protocols on L1 (Ethereum). 
 This insignificance on TR is consistent with previous findings \cite{sun2022liquidity} (see Table~\ref{tab:prior-work}, row 3). For v3 deployed on L1 and L2 blockchains, we see that TR is {\em positively} related to liquidations, i.e., with 1\% increase in liquidation, TR will increase by $0.02\%$ on L1 ($\mathrm{Liq} + \mathrm{Liq\mathord{:}Ver} = 0.00+ 0.02=0.02$), and $0.11\%$ on L2 ($\mathrm{Liq} + \mathrm{Liq\mathord{:}Lay} + \mathrm{Liq\mathord{:}Ver} = 0.00+0.09 +0.02=0.11$). This is a new finding that has not been reported before. Moreover, we also see that this positive effect is stronger for the Arbitrum-based (L2) protocols, due to a significant coefficient of $\mathrm{Liq\mathord{:}Lay}$. 

\subsection{Cross-protocol and cross-version analysis of liquidations}
\noindent
Model 3 in Table~\ref{tab:MergedTables} presents the regression on TVL for v2 and v3 versions of Aave and Compound. For Compound v2, as the baseline in Model 3, an insignificant coefficient for $\mathrm{Liq}$ was found, so there is no significant relationship between liquidations and TVL. However, for Aave v2, we find that TVL has a statistically significant {\em negative} relationship with liquidations, such that a $1\%$ increase in liquidations is associated with a $0.01\%$ {\em decrease} in TVR ($\mathrm{Liq} + \mathrm{Liq\mathord{:}Platf} = 0.00 + (-0.01) = -0.01$). In contrast, v3 protocols both show a significant {\em positive relationship} between liquidations and TVL. For Compound v3, we see that a $1\%$ increase in liquidations leads to a $0.05\%$ {\em increase} in TVL ($\mathrm{Liq} +\mathrm{Liq\mathord{:}Ver} = 0.00+ 0.05 = 0.05$); and for Aave v3, we see that a $1\%$ increase in liquidations leads to a $0.04\%$ {\em increase} in TVL ($\mathrm{Liq} + \mathrm{Liq\mathord{:}Ver} + \mathrm{Liq\mathord{:}Platf} = 0.00 + 0.05 + (-0.01) = 0.04$).
 
Model 4 in Table~\ref{tab:MergedTables} reports regression on TR for v2 and v3 versions of Aave and Compound. Significant heterogeneity across versions can be observed. For all v2 protocols, the impact of liquidation on TR is insignificant (see $\mathrm{Liq}$ and $\mathrm{Liq\mathord{:}Platf}$, respectively). However, there is a significant {\em positive} relationship between liquidation and TR for Compound v3 and Aave v3 protocols, with a $1\%$ increase in liquidations leading to a $0.06\%$ {\em increase} in TR ($\mathrm{Liq\mathord{:}Ver}=0.06$).  These results confirm that the v3 protocols have more efficient and robust liquidity risk management than the older v2 protocols. This finding is consistent with the results reported by \cite{sinyugin2023decentralized}, which showed that liquidations, in some cases, reflect a protocol's stability against risks.

\subsection{Discussion}
\noindent
Decentralised lending offers innovative opportunities for direct interaction between borrowers and lenders \cite{harvey2021defi}, yet it also introduces new risks that lack centralised governance. The liquidation process functions as a risk mitigation strategy; but its effectiveness decreases as risk levels increase and, if implemented too late, it may fail to recover the debt position \cite{sick2023economic}. However, an increased volume of liquidations does not necessarily indicate a risk for a platform; instead, sometimes, it demonstrates the protocol's ability to effectively reduce risk and protect funds within the system. Especially if the functioning of the protocol remains unaffected by significant liquidations \cite{sinyugin2023decentralized}. The results hold for our research as the regression analyses for version 3 of Aave and Compound on the Ethereum and Arbitrum blockchains exhibit a positive correlation between liquidations and both TVL and TR, reflecting the protocols' stability and health against liquidation risk. However, versions 2 of both platforms do not exhibit this positive correlation. The behavioral divergence between v2 and v3 is likely influenced by a combination of protocol-specific and market-driven factors.

Following the launch of the v3 protocols, substantial user migration was observed from v2 to v3. To smooth this shift, v3 smart contracts were embedded with integrated migration tools, advancing the user migration process. Fig.~\ref{fig:Users} shows the daily active user count for both the v2 and v3 protocols, indicating a significant transition in user engagement towards v3 protocols during the period Jan 2021 to Dec 2024. This pattern offers a plausible explanation for the increase in TVL in v3 protocols, especially after liquidation events. Instead of reinvesting in v2, borrowers choose to transition to v3, driven by its improved risk management systems and advanced functionalities. The reported increase in TVL and TR on v3 protocols---particularly on the L2 Arbitrum chain during the period of this study---can be attributed to this influx of users (see Fig.~\ref{fig:Users} and Fig.~\ref{fig:DailyDeposits}, showing high user activity and high deposit counts on Aave v3 Arbitrum). The increase in engagement from both depositors and borrowers during market stress events such as liquidations highlights the growing preference for the more resilient and user-centric design of the v3 protocols. However, the number of direct deposits and loans on the v3 protocols is significantly higher than migrated deposits and loans (Source: \href{https://dune.com/}{Dune Analytics}).

Another contributing reason to the observed increase in TVL after liquidation events, notably in v3 cross-chain lending protocols, is the rebound effect commonly observed in cryptocurrency markets. This phenomenon denotes a rapid recovery in the prices of volatile assets, such as Bitcoin and Ether, after a significant decrease. The rapid increase in asset prices following a liquidation can inflate the TVL as collateral values recover and users restore faith in market conditions. Within the framework of v3 protocols, this rebound effect can amplify TVL growth, intensifying the impact of increased user migration and continued protocol engagement.

Fig.~\ref{fig:Deposits} shows that the Ethereum blockchain in v3 lending protocols represents the highest total deposit volume in USD, indicating a substantial capital concentration. In contrast, Fig.~\ref{fig:DailyDeposits} illustrates that the Arbitrum chain experiences the highest number of individual deposit transactions. Moreover, Fig.~\ref{fig:Deposits} also confirms that the Arbitrum chain has the largest number of active users, indicating a predominantly retail-oriented user base. These findings suggest a divergence in user behaviour between chains, with large investors tending to prefer the Ethereum (L1) chain due to its robust liquidity, institutional involvement, and more integrated infrastructure within the DeFi ecosystem; while smaller retail investors primarily use Arbitrum-based (L2) protocols, likely motivated by its reduced transaction fees and faster execution. We have shown that the sensitivity of TVL and TR to liquidation events is significantly greater on the Arbitrum chain compared to Ethereum (see Table~\ref{tab:MergedTables}). This increased responsiveness illustrates the behavioural characteristics of retail investors, who are generally more sensitive to market fluctuations. These observations align with the findings of \cite{cornelli2024defi}, which indicate that retail investors are more sensitive to ETH price changes than large investors.

Furthermore, the change in the relationship with liquidations on the v2 and v3 protocols can be attributed to the varying methodologies used to execute functions such as collateralisation ratios and interest rate models, leading to significant differences in their robustness, efficiency, and resilience \cite{sinyugin2023decentralized}. In v3 protocols, liquidations occur with greater frequency; however, they are managed more effectively. Their upgraded mechanism facilitates liquidations without significant depletion of TVL and, in some cases, liquidations can even increase TVL and TR depending on the architecture, mechanism, and governance of the protocols. 


\section{Conclusion}\label{Sec:conc}
\noindent
We have investigated the effectiveness of automated risk management strategies during liquidation events and their impact on the performance and stability (measured by TVL and TR) of decentralised lending protocols. For this purpose, we selected the top lending protocols, Aave and Compound, including their old (v2) and upgraded (v3) versions deployed on L1 and L2 blockchains, to assess the effectiveness of prevalent risk management mechanisms. The results suggest that liquidation events in v3 protocols lead to increased TVL and total revenue, with a more pronounced effect on the Arbitrum (L2) blockchain compared to the Ethereum (L1) blockchain. In contrast, for the v2 protocols, liquidations have an insignificant impact on TVL and total revenue. A combination of protocol-specific and market-driven factors is likely to influence the behavioral divergence between v2 and v3. Our research has important implications for risk managers, investors, and policy makers. 

This research could be further extended by investigating the wallet-level behavior of lenders, borrowers, and liquidators during and after liquidation events to assess strategic user actions. Furthermore, an event study could be used to explore the role of specific stablecoins and volatile assets in causing systemic risk or spillover effects, after cascading liquidations or oracle failure. Another avenue of investigation is to explore the total value redeemable (TVR) as a performance measure alternative to TVL, as it has been proposed as a more accurate quantifier of the value stored within a protocol \cite{luo2024piercing}. In the longer term, our aim is to explore the effects of different risk mechanisms using agent-based simulation modelling \cite{ai4ci-position-2024-fullnames}.

\ifnum\ANON=1 {}
\else
    \section*{Acknowledgements}
    \noindent
    J. Cartlidge is supported by UK Research and Innovation (UKRI) Engineering and Physical Sciences Research Council (EPSRC) Grant Number EP/Y028392/1: AI for Collective Intelligence (AI4CI). E. Iftikhar's research is supported by PNRR (National Recovery and Resilience Plan) scholarship ex D.M. 351/2022 under National Call for PhD in Blockchain and Distributed Ledger Technology offered by University of Camerino (Administrative University) and Universita Cattolica del Sacro Cuore, Milan (Host University). This research was completed at the University of Bristol while E. Iftikhar was a visiting PhD scholar under the supervision of W. Wei and J. Cartlidge.   
\fi


\bibliographystyle{IEEEtran}
\bibliography{defi-lending}

\begin{appendices}
{
    \renewcommand{\thetable}{\Alph{section}.\arabic{table}} 
    \setcounter{table}{0}

    \section{Multicollinearity Test}\label{app:multicoll}
    \noindent
    We test multicollinearity of the variables before conducting regression, using the Variance Inflation Factor (VIF) method. These estimates of VIF fall within the range 1.1 to 2.1, as shown in Table~\ref{tab:VIF}, indicating that there is no presence of multicollinearity between the independent variables.
    
\begin{table}[H]
\centering
\caption{Variance Inflation Factor (VIF) Results} 
\label{tab:VIF}
\setlength{\tabcolsep}{2pt} 
\begin{tabular}{c cccccccc}
\toprule
\bf{Variable} & Liq & DRM & VolETH & VolGasPri & VolUSDCbor & FGI & VIX & Wd \\ 
\midrule
\bf{VIF} &1.25 & 1.13 & 1.28 & 1.61 & 1.61 & 2.11 & 1.84 & 1.36 \\ 
\bottomrule
\end{tabular}

\end{table}
    
    \section{Comprehensive Model}\label{app:FullModel}
    \noindent
    Here we have used all variable mentioned in our regression model (see Section~\ref{sec:Method-performance}) to account for cross-version, cross-chain and cross-protocol affect on protocol performance (proxied by TVL and TR). Our findings in Table~\ref{tab:FullModel} are consistent with the conclusions we made in Section~\ref{sec:Results}. The results show that there is significant heterogeneity between layers and versions, which is consistent with the results of Models 1--4. 
    
    \begin{table}[H]
\centering
\caption{Cross-Version, Cross-Chain, and Cross-Protocol Regression on TVL and Total Revenue}
\label{tab:FullModel}
\begin{tabularx}{0.6\linewidth}{l *{2}{S[table-format=+1.2, table-space-text-post=***]}}
\toprule
\textbf{Variable} & \textbf{Model 5} & \textbf{Model 6} \\
\midrule
 Liq & 0.00 & 0.00 \\
 Liq:Lay & 0.05~$^{***}$ & 0.09~$^{***}$ \\
 Liq:Ver & 0.03~$^{***}$ & 0.02~$^{*}$ \\
 Liq:Platf & -0.01~$^{*}$ & 0.00 \\
 DRM & 0.01~$^{***}$ & 0.01~$^{***}$ \\
 VolETH & -1.93~$^{**}$ & -2.23~$^{**}$ \\
 VolGasPri & -0.15~$^{***}$ & -0.19~$^{***}$ \\
 VolUSDCbor & 0.02~$^{***}$ & 0.05~$^{***}$ \\ 
 Wd & 0.40~$^{***}$ & 0.49~$^{***}$ \\
 FGI & 0.08~$^{**}$ & 0.46~$^{***}$ \\
 VIX & -0.27~$^{***}$ & -0.76~$^{***}$ \\
 Adj R-sq & 0.61 & 0.57  \\
\bottomrule
\end{tabularx}
\begin{minipage}{0.95\linewidth}
\footnotesize
\setlength{\baselineskip}{7pt} 
\vspace{1mm}
Note: This table includes estimates of two panel regression models performed on cross-chain, cross-version, cross-protocol TVL (Model 5) and TR (Model 6) as dependent variables, regressed against liquidations and relevant control variables. The models are estimated as follows:
\\

Model 5: \(
\mathrm{TVL}_{i,t}    =  \, \alpha_i
+ \beta_1 \mathrm{Liq}_{i,t}
+  \beta_2 \mathrm{Liq}_{i,t} \mathrm{Lay}_{i}
+  \beta_3 \mathrm{Liq}_{i,t} \mathrm{Ver}_{i}
 +\beta_4 \mathrm{Liq}_{i,t} \mathrm{Platf}_{i}
+ \gamma_1 \mathrm{DRM}_{i,t}
+ \gamma_2  \mathrm{VolETH}_{t} 
+ \gamma_3  \mathrm{VolGasPri}_{i,t}\nonumber
+\gamma_4 \mathrm{VolUSDCbor}_{i,t}
+ \gamma_5 \mathrm{Wd}_{i,t} 
+ \gamma_6 \mathrm{FGI}_{t}
+ \gamma_7 \mathrm{VIX}_{t} 
+  \varepsilon_{i,t}\) \\

Model 6: \(
\mathrm{TR}_{i,t}    =  \, \alpha_i
+ \beta_1 \mathrm{Liq}_{i,t}
+  \beta_2 \mathrm{Liq}_{i,t} \mathrm{Lay}_{i}
+  \beta_3 \mathrm{Liq}_{i,t} \mathrm{Ver}_{i} +\beta_4 \mathrm{Liq}_{i,t} \mathrm{Platf}_{i}
+ \gamma_1 \mathrm{DRM}_{i,t}
+ \gamma_2  \mathrm{VolETH}_{t} 
+ \gamma_3  \mathrm{VolGasPri}_{i,t}
+\gamma_4 \mathrm{VolUSDCbor}_{i,t}
+ \gamma_5 \mathrm{Wd}_{i,t} 
+ \gamma_6 \mathrm{FGI}_{t} 
+ \gamma_7 \mathrm{VIX}_{t} 
+  \varepsilon_{i,t}\) \\

Asterisks indicate statistical significance of estimates; with $^{*}$ indicating $p < 0.10$, $^{**}$ indicating $p < 0.05$, and $^{***}$ indicating $p < 0.01$.

\end{minipage}

\end{table}
}
\end{appendices}

\end{document}